\documentclass{blois}

\def\be{\begin{equation}}
\def\ee{\end{equation}}
\def\bea{\begin{eqnarray}}
\def\eea{\end{eqnarray}}

\newcommand{\Photo}{}

\begin{document}
\vspace*{4cm}
\title{THEORETICAL PERSPECTIVES ON FLAVOUR PHYSICS}
       
\author{R. FLEISCHER}
\address{Nikhef, Science Park 105, 1098 XG Amsterdam, Netherlands and Department of Physics \\ 
and Astronomy, Vrije Universiteit Amsterdam, 1081 HV Amsterdam, Netherlands}

\maketitle\abstracts{Flavour physics and CP violation play a key role for the testing of the Standard Model and search for New Physics at the high-precision frontier. After a compact discussion of the current status of (quark)-flavour physics, I will illustrate theoretical prospects for benchmark processes in the sector of decays of $B$ mesons, ranging from precision determinations of the CP-violating phases of neutral $B$-meson mixing over non-leptonic decays governed by penguin topologies and channels arising only from tree diagrams to rare loop-induced decays into final states with leptons. Exciting opportunities for theorists and experimentalists arise in the era of the HL-LHC and Belle II and beyond at the FCC-ee.}

\section{Setting the Stage}
In the history of the Standard Model (SM), flavour physics and CP violation were key players and have continued to progress over recent decades. Within the SM, quark-flavour mixing is described by the Cabibbo--Kobayashi--Maskawa (CKM) matrix, connecting the quark flavour states with their mass eigenstates. The CKM matrix is encoded in weak decays of $K$, $D$ and $B$ mesons, offering powerful probes for testing this framework and to search for footprints of New Physics (NP). The key challenge in this endeavour is related to strong interactions since the theory is formulated in terms of quarks while experiments use their bound states.  In calculations of the relevant transition amplitudes, we encounter process-dependent, non-perturbative ``hadronic" parameters. 

In order to deal with the huge hierarchy of energy scales arising in these studies, ranging from NP scales $\Lambda_{\rm NP}$ 
in the TeV regime -- and possibly far above -- over the electroweak scale to long-distance dynamics at $\Lambda_{\rm QCD}\sim 10^{-4} \,\mbox{TeV}$, we have powerful theoretical tools given by effective field theories: Heavy degrees of freedom, such as NP particles, the top quark and the $Z$, $W$ bosons of the SM are ``integrated out” from appearing explicitly and are described by short-distance functions. Calculations of QCD corrections were performed in systematic ways utilising renormalisation group techniques.\cite{AJB} The effective 
field theory framework provides also an important setting for NP studies, where Standard Model Effective Theories (SMEFT) are very popular.

We have indications that the SM is not complete, such as the non-vanishing neutrino masses, the baryon asymmetry of the Universe and the long-standing problem of dark matter. Moreover, we don't know the origin of the structures of the SM and its hierarchies, which may well be associated with New Physics (NP). Popular extensions of the SM, such as models with extra $Z'$ bosons or leptoquarks, usually involve also new sources of flavour and CP violation. 

How to search for physics beyond the SM? Most obviously, we may try to produce new particles at colliders at the high-energy frontier and study their decays in general purpose detectors, where the current key player is the Large Hadron Collider (LHC) with the ATLAS and CMS experiments. On the other hand, we may also search for indirect signals of NP through virtual quantum effects at the high-precision frontier, where the current experimental key players for quark-flavour physics are the LHCb and Belle II experiments.\cite{Amhis}

Let us have a closer look at key flavour physics probes: Historically, the kaon system led to the discovery of indirect and direct CP violation in 1964 and 1999, respectively, through $K\to\pi\pi$ decays. Since decades, the rare kaon decays $K^+\to \pi^+\nu\bar\nu$ and $K_{\rm L}\to \pi^0 \nu\bar\nu$, which are theoretically very clean SM probes, are the main interest in kaon physics. It is very important to continue the exploration of kaon physics.\cite{kaon} The $D$-meson system is characterised by small CP violation and $D^0$--$\bar D^0$ mixing in the SM due to the structure of the CKM matrix and down-type quarks running in penguin and box topologies. In weak $D$ decays, we have generally significant non-factorisable and $SU(3)$-breaking effects, providing interesting probes for QCD and long-distance dynamics. CP violation was established through a difference of CP asymmetries in $D^0\to K^+K^-, \pi^+\pi^-$ decays. The question whether the observed effect can be accommodated in the SM or requires NP has yet to be answered. Exciting perspectives arise in studies of CP violation and rare $D$ decays (for a recent review, see Ref.~\cite{D-rev}). The $B$-meson system will be the focus of the following discussion. Moreover, probes for Lepton Flavour Violation (leading to processes such as $\mu\to e\gamma$, ...), electric dipole moments, etc., offer interesting complementary processes to search for NP effects. 

Where do we stand? The physics explorations at the LHC have led to the discovery of the Higgs particle, but no SM deviations have so far been observed at ATLAS and CMS. We have also no solid evidence for NP in the flavour sector, although some ``puzzles" with respect to the SM -- also at Belle II -- lead to excitement. The general implications for the structure of NP are that we have a large NP scale, which would be bad news for the direct searches at the LHC or/and we have symmetries at work which prevent large effects from physics 
beyond the SM in flavour-changing neutral current (FCNC) processes and the flavour sector. Much more is still to come but we have to prepare to deal with smallish and challenging NP effects. Exciting years for flavour physics are ahead of us: It will be very interesting to see which insights the analyses and theoretical interpretation of the LHC run 3 data in the next years will reveal. In the period until around 2040, we will have the full exploitation of Belle II and the High Luminosity Large Hadron Collider (HL-LHC) with the LHCb Upgrade II, keeping experimentalists and theorists very busy. 

In the era after the HL-LHC, exciting new prospects are offered by the FCC-ee,\cite{MW-21} 
raising questions about how particle physics will look like around the year 2050: Will discrepancies with respect to the SM have been established? Will new particles have been observed? Will new sources of CP violation have been found? What about the progress in theoretical particle physics? Within the 2026 update of the European Strategy for Particle Physics, various future scenarios were discussed, also with respect to flavour physics.\cite{Physics-Briefing-Book} Illustrations of the expected progress for the determination of the apex of the Unitarity Triangle (UT) of the CKM matrix are particularly impressive,  with projections to the FCC-ee era. Interestingly, limitations are given by the CKM matrix elements $|V_{cb}|$ and $|V_{ub}|$, which are 
unfortunately still affected by long-standing tensions between inclusive and exclusive determinations.\cite{UT-NP} The FCC-ee will also have an important  impact on this topic, such as through the determination of $|V_{cb}|$ through on-shell $W$ decays into charmed and 
beauty jets.

A central element in the determination of the UT is given by CP violation in $B$ decays, where non-leptonic channels play the key role due to  interference phenomena which arise in such processes.\cite{RF-CP-rev} However, studies of CP violation go far beyond the UT and offer very powerful probes for testing the SM and revealing possible new source of CP violation. In this presentation, I will focus on CP-violating phenomena in $B$ decays and present a selection of processes which -- from a theoretical point of view -- offer exciting perspectives for flavour physics.

\boldmath
\section{Precision Measurements of the $B^0_q$--$\bar B^0_q$ Mixing Phases}\label{sec:1}
\unboldmath
The phenomenon of $B^0_q$--$\bar B^0_q$ mixing ($q=d,s$) arises from box diagram topologies in the SM and is very sensitive to contributions from NP, which may enter through new particles in the loop topologies or even at the tree level as, for instance, in models with extra $Z'$ bosons. In particular the associated CP-violating phases $\phi_q$ offer an outstanding opportunity to 
search for new sources of CP violation at the high-precision frontier. These phases can be written as
\begin{equation}\label{phi_q}
\phi_d=2\beta+\phi_d^{\rm NP}, \quad \phi_s=-2\lambda^2\eta +\phi_s^{\rm NP},
\end{equation}
where $\beta$ is the usual angle of the UT, $\lambda\equiv|V_{us}|\approx 0.22$ and $\eta$ are CKM parameters, and the $\phi_q^{\rm NP}$ denote CP-violating NP phases. In order to reveal such NP effects through measurements, it is crucial to have a critical look at the corresponding theoretical SM analyses and the underlying assumptions, where the challenge is generally given by hadronic effects arising from strong interactions. The goal is to match the experimental with the theoretical uncertainties.

The benchmark decays are $B^0_d\to J/\psi K_{\rm S}$ and $B_s^0\to J/\psi \phi$, which allow measurements of the CP-violating phases $\phi_d$ and $\phi_s$, respectively. The experimental prospects for the precision of these determinations up to the era of the 
FCC-ee are very impressive, as illustrated in the Physics Briefing Book.\cite{Physics-Briefing-Book} 
The theoretical uncertainties in these analyses arise from doubly Cabibbo-suppressed penguin
contributions.\cite{RF-97,RF-99} The increasing experimental precision requires the control of these effects to reveal possible new sources of CP violation. This topic has received long-standing interest in the particle physics community.\cite{RF-CP-rev} The hadronic corrections to the measured values of the phases $\phi_q$ cannot be reliably calculated within QCD due to their non-perturbative character. 
However, nature offers channels to control these effects with the help of flavour symmetries of strong interactions and experimental data.

Let us have a look at the ``golden" decay $B^0_d\to J/\psi K_{\rm S}$. Within the SM, its decay amplitude can be written as follows:\cite{RF-99}
\begin{equation}\label{ampl-1}
A(B_d^0\to J/\psi\, K_{\rm S})=\left(1-\lambda^2/2\right){\cal A'}
\left(1+\epsilon\, a'e^{i\theta'}e^{i\gamma}\right),
\end{equation}
where ${\cal A'}$ and $a'e^{i\theta'}$ are CP-conserving hadronic parameters. Whereas the former is governed by colour-suppressed tree topologies, the latter measures the ratio of penguin to tree contributions, and enters with the doubly Cabibbo-suppressed parameter 
$\epsilon\equiv\lambda^2/(1-\lambda^2)\sim 0.05$ and the CP-violating phase $\gamma$, which is the usual angle of the UT. 
Measuring time-dependent CP-violating rate asymmetries, the phase
\begin{equation}
\phi_d^{\rm eff}\equiv\phi_d+\Delta\phi_d
\end{equation}
can be extracted from the data, where $\Delta\phi_d$ is a hadronic phase shift:  
\begin{equation}
\sin\Delta\phi_d \propto 2 \epsilon a'\cos\theta' \sin\gamma+\epsilon^2a'^2, \quad
\cos\Delta\phi_d \propto 1+ 2 \epsilon a'\cos\theta' \cos\gamma+\epsilon^2a'^2
\cos2\gamma.
\end{equation}
If we neglect the doubly Cabibbo-suppressed penguin parameter $\epsilon a'$, the CP-violating phase $\phi_d^{\rm eff}$ simply reduces to $\phi_d$, which is given by $2\beta$ in the SM, as can be seen in Eq.~(\ref{phi_q}). 

In order to determine the hadronic phase shift, we may utilise CP violation in the $B^0_s\to J/\psi K_{\rm S}$ channel, which is related to 
$B^0_d\to J/\psi K_{\rm S}$ through the $U$-spin symmetry of strong interactions, i.e.\ the interchange of down and strange quarks, 
and has the following SM amplitude:\cite{RF-99}
\begin{equation}
A(B^0_s\to J/\psi K_{\rm S})=-\lambda{\cal A}\left(1-a e^{i\theta}e^{i\gamma}\right),
\end{equation}
where we have the $U$-spin relations $a=a'$ and ${\cal A}={\cal A'}$. In contrast to Eq.~(\ref{ampl-1}), $a e^{i\theta}$ does not enter with the doubly Cabibbo-suppressed parameter $\epsilon$. Measuring the time-dependent CP asymmetries in $B^0_s\to J/\psi K_{\rm S}$, the penguin parameters $a$ and $\theta$ can be determined in a clean way. Using then the $U$-spin symmetry, we can relate them to their counterparts $a'$ and $\theta'$, allowing us to include the hadronic effects in the determination of $\phi_d$ from the 
$B^0_d\to J/\psi K_{\rm S}$ decay.\cite{RF-99}

A similar situation arises in the determination of $\phi_s$ from CP violation in $B^0_s\to J/\psi \phi$. Since we have two vector mesons in the final state, an angular analysis is needed in this case. Various control channels and measurements of their CP asymmetries can be 
added,\cite{pen-25} resulting in a complex strategy to include the doubly Cabibbo-suppressed penguin effects in the determinations of 
the mixing phases, yielding $\phi_d=(45.7^{+1.1}_{-1.0})^\circ$ and $\phi_s=(-3.72^{+1.03}_{-0.97})^\circ$ from the current data. For future projections, see the detailed discussion in Ref.~\cite{pen-25}.

Decays of the kind $B^0_{d}\to J/\psi K_{\rm S}$ and $B^0_s\to J/\psi \phi$ should be fully exploited at the HL-LHC and Belle II. The control of the hadronic penguin effects will be crucial and equally important to improving precision on these golden modes: Will we already
reveal a SM discrepancy for $\phi_{s(d)}$ in the next years? The $U$-spin partner $B^0_s\to J/\psi K_{\rm S}$ of
the $B^0_d\to J/\psi K_{\rm S}$ channel offers the cleanest control mode; CMS has recently reported the currently best measurement of CP violation in this decay.\cite{CMS-BspsiKS} It will be interesting to compare with $B^0_d\to J/\psi \pi^0$ and further control channels, and to study time-dependent CP violation in $B^0_d\to J/\psi \rho^0$ angular analyses, providing also valuable insights into hadron dynamics. 
The experimental prospects motivates theorists to develop new strategies. Beyond the era of the HL-LHC and Belle II, which precision can be achieved at the FCC-ee? The flavour physics potential is currently explored within a workshop series at CERN.\cite{FlavourFCC-Workshop}

\boldmath
\section{Non-Leptonic $B$ Decays Governed by Penguin Topologies}
\unboldmath
The $B\to \pi K$ system provides a powerful laboratory for testing the SM. Interestingly, these decays are governed by QCD penguin topologies. In the case of the decays $B^0_d\to \pi^-K^+$ and $B^+\to \pi^+K^0$, electroweak (EW) penguins are colour-suppressed and play a minor role. On the other hand, EW penguins are colour-allowed in the $B^0_d\to\pi^0K^0$ and $B^+\to\pi^0K^+$ channels, resulting in sizeable contributions, competing even with colour-allowed tree amplitudes. The decay $B^0_d\to \pi^0K_{\rm S}$ stands out as it is the only of the $B\to\pi K$ channels showing mixing-induced CP violation:
\begin{equation}\label{CP-asym-1}
	\frac{\Gamma(\bar{B}_d^0(t) \rightarrow \pi^0K_{\rm S}) - 
\Gamma(B_d^0(t) \rightarrow \pi^0K_{\rm S})}{\Gamma(\bar{B}_d^0(t) \rightarrow \pi^0K_{\rm S}) + 
\Gamma(B_d^0(t) \rightarrow \pi^0K_{\rm S})} 
= A^{\pi^0K_{\rm S}}_{\rm CP} \cos(\Delta M_dt) + S^{\pi^0K_{\rm S}}_{\rm CP}\sin(\Delta M_dt).
\end{equation}
Here the time dependence arises from the $B^0_d$--$\bar B^0_d$ oscillations and $A^{\pi^0K_{\rm S}}_{\rm CP}$ and 
$S^{\pi^0K_{\rm S}}_{\rm CP}$ describe direct and mixing-induced CP violation, respectively. The CP asymmetries satisfy the relation
\begin{equation} 
	S^{\pi^0K_{\rm S}}_{\rm CP} = \sin(\phi_d - \phi_{00})\sqrt{1- (A^{\pi^0K_{\rm S}}_{\rm CP})^2},
\end{equation}
where $\phi_d$ is again the $B^0_d$--$\bar B^0_d$ mixing phase and $\phi_{00} \equiv \rm{arg}(\bar{A}_{00} A^*_{00})$ the angle between the decay amplitude $A_{00} \equiv A(B_d^0 \to \pi^0 K^0)$ and its CP-conjugate $\bar{A}_{00}$. In order to determine $\phi_{00}$, we may utilise amplitude triangles which can be fixed through branching ratios and direct CP asymmetries of the neutral $B\to\pi K$ modes.\cite{FJPZ} They are related to the following isospin relation:
\begin{equation} 
	\sqrt{2} A(B^0_d \to \pi^0 K^0) + A(B^0_d \to \pi^- K^+)   \equiv 3A_{3/2},
\end{equation}
where the $I=3/2$ amplitude $A_{3/2}$ depends on the EW penguin contributions and can be determined with a minimal $SU(3)$ flavour symmetry input through the $B^+\to\pi^+\pi^0$ branching ratio. Following these lines, a SM correlation between the direct and mixing-induced CP asymmetries of the $B^0_d\to \pi^0K_{\rm S}$ decay can be calculated, which shows an intriguing tension with experimental 
data.\cite{FJPZ,FJMV-18,EM-Moriond} It will be interesting to reveal the origin of this pattern, which may be due to a modified EW penguin sector with new sources of CP violation originating from beyond the SM.

Sophisticated strategies were developed utilising the whole $B\to\pi K$ system as well as $B \to \pi\pi$ decays.\cite{EM-Moriond} 
Recently, we have seen an interesting new measurements of CP violation in $B^0\to\pi^0\pi^0$ by the Belle II collaboration.\cite{Belle-II-pi0pi0} Will we already establish SM discrepancies in the next years? As noted, EW penguins offer an exciting possibility for NP to enter: Examples are models with extra $Z'$ bosons, playing also an important role as a NP portal in rare $B$ decays discussed in 
Section~\ref{sec:rare}.

Another powerful non-leptonic $B$ decay governed by QCD penguins is the $B^0_s\to K^+K^-$ channel, which can be complemented through $B^0_d\to \pi^+\pi^-$ with the help of the $U$-spin symmetry to extract the UT angle $\gamma$ from the corresponding CP asymmetries.\cite{RF-BsKK-99} Current data yield $\gamma=(65^{+11}_{-7})^\circ$, in excellent agreement with the SM and 
other determinations.\cite{FJV}  Further interesting probes of the SM and isospin-breaking NP contributions are provided by $B^0_d\to \phi K_{\rm S}$, $B^\pm\to \phi K^\pm$ decays.\cite{FM-01,FGV}

\boldmath
\section{Non-Leptonic $B$ Decays Originating from Tree Topologies}\label{sec:tree}
\unboldmath
A prime example of this decay class is given by the $B^0_s\to D_s^\mp K^\pm$ system, offering yet another exciting way to probe the SM through CP-violating phenomena.\cite{ADK-1992,RF-BsDsK-2003} These channels originate from colour-allowed tree topologies. Since both the $B^0_s$ and the $\bar B^0_s$ mesons may decay into the $D_s^+K^-$ final state and its CP conjugate, $B^0_s$--$\bar B^0_s$ mixing generates interference effects, which result in mixing-induced CP violation and the following time-dependent rate 
asymmetry:\cite{RF-BsDsK-2003}
\begin{equation} 
	\frac{\Gamma(B^0_s(t)\to D_s^{+} K^-) - \Gamma(\bar{B}^0_s(t)\to D_s^{+} K^-) }
	{\Gamma(B^0_s(t)\to D_s^{+} K^-) + \Gamma(\bar{B}^0_s(t)\to D_s^{+} K^-) }  
= \frac{{C}\,\cos(\Delta M_s\,t) + {S}\,\sin(\Delta M_s\,t)}
	{\cosh(y_s\,t/\tau_{B_s}) + {\cal A}_{\Delta\Gamma}\,\sinh(y_s\,t/\tau_{B_s})},
\end{equation}
which is analogous to Eq.~(\ref{CP-asym-1}) for a final CP eigenstate, and takes also the effects of the sizeable decay width difference of the $B_s$ system through $y_s\equiv \Delta\Gamma_s/(2\,\Gamma_s)$ into account. We have
\begin{equation}
 C=\frac{1-|\xi|^2}{1+|\xi|^2},  \quad S= \frac{2\,{\rm Im}{\,\xi}}{1 + |\xi|^2}, \quad 
 \mathcal{A}_{\Delta \Gamma}=\frac{2\,{\rm Re}\,\xi}{1+|\xi|^2} ,
\end{equation}
and similar structures for the CP-conjugate final state $D_s^-K^+$, with $\overline{C}$, $ \overline{S}$, 
$\overline{{\cal A}}_{\Delta\Gamma}$ and ${\bar{\xi}} $. The observables $\xi$, ${\bar{\xi}} $ describe the interference effects between the decay paths and can be written as 
\begin{equation}
  \xi=  - e^{-i(\phi_s + \gamma)}  \left[\frac{1}{x_s e^{i \delta_s}} \right], \quad 
  {\bar{\xi}}   = - e^{-i(\phi_s + \gamma)} \left[ x_s e^{i \delta_s} \right],
\end{equation}
where $x_s e^{i \delta_s}$ is a non-perturbative hadronic parameter characterizing the ratio of decay amplitudes. Using the time-dependent rate asymmetries, $ \xi$ and ${\bar{\xi}}$ can be extracted, allowing us to determine 
\begin{equation}\label{prod-xi}
{\xi} \times \bar{\xi}= e^{-i2( \phi_s + \gamma)}.
\end{equation}
Since the hadronic parameters cancel in this product,\cite{RF-BsDsK-2003} it gives a theoretically clean determination of 
$\phi_s+\gamma$. With $\phi_s$ determined as discussed in Section~\ref{sec:1}, the UT angle $\gamma$ can be extracted as well. 

LHCb has performed pioneering measurements of the $B^0_s\to D_s^\mp K^\pm$ system, reporting the result $\gamma=(128^{+17}_{-22})^\circ$ in 2017, which should be compared with the SM value $\gamma\sim70^\circ$. In 2024, the result $\gamma=(81^{+12}_{-11})^\circ$ was announced.\cite{LHCb-BsDsK-2024} It will be exciting to monitor how these analyses will evolve with future data.
Interestingly, the puzzling $\gamma$ patterns are complemented by that for a parameter $a_{\rm 1 \, eff }^{D_s K}$, which characterises the factorisation of hadronic matrix elements for colour-allowed tree decays such as the $\bar B^0_s\to D_s^+ K^-$ mode.\cite{FM-BsDsK-1,FM-BsDsK-2} The cleanest way of extracting this quantity from the data is given by the following ratio:
\begin{equation}
  R_{D_s^{+}K^{-}}\equiv\frac{\mathcal{B}(\bar{B}^0_s \rightarrow D_s^{+}K^{-})_{\rm th}}{{\mathrm{d}\mathcal{B}\left(\bar{B}^0_s \rightarrow D_s^{+}\ell^{-} \bar{\nu}_{\ell} \right)/{\mathrm{d}q^2}}|_{q^2=m_{K}^2}} = 
  6 \pi^2 f_{K}^2 |V_{us}|^2 |a_{\rm 1 \, eff }^{D_s K}|^2  X_{D_s K},
\end{equation}
where $f_K$ is the kaon decay constant, $|V_{cb}|$ cancels and hadronic form factors entering $X_{D_s K}$ play a very minor role. 
The experimental value, corrected also for non-factorisable exchange contributions, is found surprisingly below the theory prediction. Interestingly, a similar pattern arises for decays with similar dynamics,\cite{FM-BsDsK-1,FM-BsDsK-2}  reaching $4.8\sigma$ for $\bar B^0_d\to D_d^+K^-$. This mode and its counterparts are benchmark decays for the application of QCD factorisation. How could NP enter? Contributions to $B^0_s$--$\bar B^0_s$ mixing would affect the mixing parameters and may lead to a NP shift of $\phi_s$. However, these effects are included when using the value of $\phi_s$ determined as discussed in Section~\ref{sec:1}. On the other hand, NP contributions entering at the decay amplitude level would affect the $\gamma$ measurement as well as the $a_1$ values, resulting in an interplay between CP violation and branching ratio measurements.\cite{FM-BsDsK-1,FM-BsDsK-2} 
Detailed analyses were performed within specific NP scenarios,\cite{IK,BGM} and general NP effect in tree decays were also considered.\cite{BLTW} The analysis of CP violation in the $B^0_s\to D_s^\mp K^\pm$ system was generalised to the presence of CP-violating NP: 
\begin{equation}
\xi \times \bar{\xi}  = \sqrt{1-2\left[\frac{C+\bar{C}}{\left(1+C\right)\left(1+\bar{C}\right)}
\right]}e^{-i\left[2 (\phi_s +\gamma_{\rm eff})\right]},
\end{equation}
where $\gamma_{\rm eff}$ is an effective angle with a NP phase shift.\cite{FM-BsDsK-1,FM-BsDsK-2} In the SM, it reduces to
Eq.~(\ref{prod-xi}). 

In the future, it would be desirable to have most precise individual $\gamma$ measurements with highest precision. Will tensions arise? 
Concerning the puzzling patterns in the branching ratios of $B^0_s\to D_s^\mp K^\pm$ and similar modes, the question is whether they 
originate from effects of QCD or NP. How will the interplay with CP violation studies and $\gamma$ determinations, which are 
theoretically clean, evolve in the future? Will we get links to possible NP effects in the charged current sector? 
These decays offer exciting topics for theorists and experimentalists.

\boldmath
\section{Rare $B$ Decays into Final States with Leptons}\label{sec:rare}
\unboldmath
A key example of this decay class is given by $B^0_q\to\ell^+\ell^-$ modes ($q=d,s$). In the SM, these channels arise through FCNC
 processes at the loop level. Moreover, the decays are helicity suppressed, which is reflected by branching ratios 
proportional to $m_\ell^2$. These modes are hence strongly suppressed in the SM. Moreover, the hadronic sector is very simple and described by the $B_q$ decay constant $f_{B_q}$, which can be impressively calculated with lattice QCD. Consequently, 
$B^0_q\to\ell^+\ell^-$ belong to the cleanest rare $B$-meson decays,\cite{BGHMSS} and even QED corrections were calculated.\cite{BBS}  These channels show high sensitivity to physics from beyond the SM. A particularly interesting feature is that new (pseudo)-scalars may lift the helicity suppression present in the SM. 

Concerning the status of the $B^0_q\to\ell^+\ell^-$ decays, only $B^0_s\to\mu^+\mu^-$ has been observed at the LHC so far with a branching ratio at the $3.5\times 10^{-9}$ level. In the case of $B^0_s\to\tau^+\tau^-$, due to the large $\tau$ mass, the helicity suppression is numerically not very efficient, whereas the $\tau$ reconstruction is very challenging at the LHC, resulting in first limits. In the case of the $B^0_q\to e^+ e^-$ modes, the helicity suppression in the SM is enormous but could be lifted through NP effects.
Consequently, it is important to search for these modes. Their observation would be a spectacular NP signal.\cite{Bsee-paper}

Interestingly, $B^0_s$--$\bar B^0_s$ mixing has also an impact on the $B^0_s\to \mu^+\mu^-$ channel, generating interference effects similar to those discussed in Section~\ref{sec:tree}, which lead to CP-violating effects and subtleties for the comparison of the measured branching ratio with the theoretical prediction:\cite{Bsmumu-ADG} 
\begin{equation}
    \mathcal{B} (B_s \to \mu^+ \mu^-)_{\rm theo} = \left[\frac{1 - y_s^2}{1 + \mathcal{A}_{\Delta\Gamma_s} y_s}\right] \overline{\mathcal{B}} (B_s \to \mu^+ \mu^-) .
\end{equation}
Here the ``experimental" branching ratio $\overline{\mathcal{B}} (B_s \to \mu^+ \mu^-)$ corresponds to the untagged, time-integrated decay rate, while the ``theoretical" counterpart describes the case where mixing effects are not included. The observable $\mathcal{A}_{\Delta\Gamma_s}$ can be accessed through the untagged effective lifetime
\begin{equation}
\tau_{\mu\mu} \equiv \frac{\int_0^\infty t\,\langle \Gamma(B_s(t)\to \mu^+\mu^-)\rangle\, dt}
	{\int_0^\infty \langle \Gamma(B_s(t)\to \mu^+\mu^-)\rangle\, dt} .
\end{equation}
We may convert the $B_s\to\mu^+\mu^-$ branching ratio data into a circular region in the $|P_{\mu\mu}|$--$|S_{\mu\mu}|$ 
plane, where $P_{\ell\ell}$ and $S_{\ell\ell}$ involve short-distance Wilson coefficients and enter the $\bar B^0_q\to\ell^+\ell^-$ decay 
amplitudes, describing pseudo-scalar and scalar contributions, respectively:\cite{Bsmumu-ADG} 
\begin{equation}
P_{\ell\ell}\equiv |P_{\ell\ell}|e^{i\varphi_P^{\ell\ell}}\equiv\frac{C_{10}^{\ell\ell}-C_{10}^{\ell\ell'}}{C_{10}^{\rm SM}}+\frac{M_{B_s}^2}{2\,{m_\ell}}
\left(\frac{m_b}{m_b+m_s}\right)\left[\frac{C^{\ell\ell}_P-C_P^{\ell\ell'}}{C_{10}^{\rm SM}}\right]
{\,\stackrel{\rm SM}{\longrightarrow}\, 1}
\end{equation}
\begin{equation}
S_{\ell\ell}\equiv |S_{\ell\ell}|e^{i\varphi_S^{\ell\ell}}\equiv\sqrt{1-4\frac{m_\ell^2}{M_{B_s}^2}}
\frac{M_{B_s} ^2}{2\,{m_\ell}}\left(\frac{m_b}{m_b+m_s}\right)
\left[\frac{C_S^{\ell\ell}-C_S^{\ell\ell'}}{C_{10}^{\rm SM}}\right]
{\,\stackrel{\rm SM}{\longrightarrow}\, 0.}
\end{equation}
The current experimental constraint includes the SM point within the uncertainties. However, it leaves a large allowed circular band, extending into NP space. In order to determine where we sit in this region, we may utilise the theoretically clean $\mathcal{A}_{\Delta\Gamma_s}$ observable.\cite{Bsmumu-ADG,Bsmumu-CPV}  We have pioneering measurements by the ATLAS, CMS and 
LHCb collaborations. Unfortunately, they are affected by too large uncertainties to have a constraining effect in the $|P_{\mu\mu}|$--$|S_{\mu\mu}|$ plane. It will be very important to obtain much more precise measurements of $\tau_{\mu\mu}$ and the associated 
$\mathcal{A}_{\Delta\Gamma_s}$.

The rare decay $B^0_s\to\mu^+\mu^-$ allows also explorations of CP violation to probe NP effects. A key example is the time-dependent 
asymmetry of the helicity-averaged decay rates:
 \begin{equation}
 \frac{\Gamma(B^0_s(t) \to \mu^+ \mu^-) - \Gamma(\bar B^0_s(t) \to \mu^+ \mu^-)}{\Gamma(B^0_s(t) \to \mu^+ \mu^-) + \Gamma(\bar B^0_s(t) \to \mu^+ \mu^-)} = \frac{{\cal S}_{\mu\mu} \sin (\Delta M_s t)}{\cosh (y_s t/\tau_{B_s}) + \mathcal{A}_{\Delta\Gamma_s} \sinh (y_s t/\tau_{B_s})},
\end{equation}
where the mixing-induced CP asymmetry ${\cal S}_{\mu\mu}$ is theoretically clean. Unfortunately, no measurements of this observable
are available, thereby leaving exciting opportunities. Interestingly, the currently allowed corridor of short-distance function governing $B^0_s\to\mu^+\mu^-$ following from the measured branching ratio may result in very constrained regions in the $\mathcal{A}_{\Delta\Gamma_s}$--${\cal S}_{\mu\mu} $ plane for real pseudo-scalar and scalar NP coefficients.\cite{Bsmumu-CPV} Should future measurements fall outside these regions, we would get immediate evidence for new sources of CP violation in this sector. Further strategies for studying CP in $B^0_s\to\mu^+\mu^-$ were proposed in the literature, including the calculation of correlations of CP-violating NP parameters and the application of 
SMEFT relations.\cite{CPV-rare-B}

The analyses of CP violation in $B^0_s\to\mu^+\mu^-$ require a short-distance Wilson coefficient $\mathcal{C}_{10}^-$ as input. This quantity as well as generally complex Wilson coefficients $ \mathcal{C}_{9}^- $, $ \mathcal{C}_{9}^+ $ as well as $\mathcal{C}_{10}^+$ can be determined through CP-violating asymmetries in $B_d^0\to K_{\rm S}\mu^+\mu^-$, $B\to K^*\mu^+\mu^-$ and $B^0_s\to \phi \mu^+\mu^-$ decays.\cite{RF-BVellell-pap} CP violation in these channels is a very ``complex" topic.\cite{BHP,DGNBV,BFKS}

CP violation offers also new probes for testing the violation of Lepton Flavour Universality (LFU) through the comparison of 
CP asymmetries in $B\to K^{(*)} \mu^+\mu^-$ and $B\to K^{(*)} e^+e^-$ decays. 
Interestingly, current data for the LFU ratios $R_{K^{(*)}}\sim1$ 
still leave significant space for such effects.\cite{RK-CPV-23} The LHCb collaboration has recently announced the first 
measurement of time-dependent CP violation in $B_d^0\to K_{\rm S}\mu^+\mu^-$, which is an important first step into a new territory.\cite{LHCb-CP-rare-2026} 

There are further exciting rare processes: Decays of the kind $B\to \pi\ell^+\ell^-$, $B\to \rho\ell^+\ell^-$, which originate from $\bar b \to d \ell^+\ell^-$ quark-level transitions in contrast to the $\bar b \to s \ell^+\ell^-$ modes discussed above, as well as $B\to K^{(*)}\nu\bar\nu$ decays and rare $B$ decays with $\tau$ leptons in the final states.

\section{Concluding Remarks}
Flavour physics continues to offer fantastic opportunities for testing the SM and revealing possible signals of NP at the high-precision frontier. In the discussion above, I have just sketched a selection of processes and topics. CP violation plays an outstanding role in this endeavour -- precision really matters! Exciting opportunities arise for theorists and experimentalists in the era of the HL-LHC and Belle II and beyond at the FCC-ee. Will we finally establish New Physics?

\section*{Acknowledgments}
I am very grateful to my students and collaborators for our flavour physics adventures over the recent years. Research presented in this 
contribution was supported by the Netherlands Organisation for Scientific Research (NWO). I would like to thank Christophe Grojean and Jacques Dumarchez and their co-organisers for the invitation and organising the wonderful conference in such a spectacular setting. It was a very inspiring meeting.

\end{document}